# Investigation of resonant polarization radiation of relativistic electrons in gratings at small angles

A.N. Aleinik, O.V. Chefonov, B.N.Kalinin, G.A. Naumenko[*], A.P. Potylitsyn,

G.A.Saruev, A.F. Sharafutdinov

Institute for Nuclear Physics, pr. Lenina 2a, 634050, Tomsk, Russia.

___________________________________________________________________________

**Abstract**

The Resonant Optical Polarization Radiation (ROPR) in the Smith-Purcell geometry and the one from the inclined grating at the Tomsk synchrotron and 6-MeV microtron have been investigated. The polarization radiation was observed at $4.2^o$ from the 200 MeV electron beam and at $5^o$ from the 6.2 MeV electron beam. Two methods of measurement of ROPR maxima in these two cases have been used. In the first case (the experiment on synchrotron) we have fixed the wavelength of radiation using an optical filter; the orientation dependence of this radiation was measured. In this dependence we have observed two peaks of radiation from electrons in gold foil grating of 0.1 mm period. The first large peak is a zeroth order peak in direction of specular reflection, and the second one is the 1-st order peak of resonant polarization radiation. In the experiment on microtron the spectra of ROPR from aluminum foil strip grating of 0.2 mm period in the Smith-Purcell geometry were measured, and the peak of the 1-st order Smith-Purcell radiation in these spectra was observed. The comparison of data obtained with the simulation results has been performed.

[*] Corresponding author. E-mail Naumenko@npi.tpu.ru

___________________________________________________________________________

1. **Introduction**

   In the late 90s the studies of radiation in conducting targets effected by relativistic electron field [1-4] became more active. One of the reasons of such activization is the need for noninvasive diagnostics of charged particle beams. It may be noted that the field of investigations moved to the optical radiation range, because in this field there are simple technical means for measuring not only angular but also spectral distribution of radiation, which may considerably increase the possibilities of beam diagnostics. Radiation from periodic targets is attractive because of the presence of resonances in direction, which is determined by the structure period of the target and the wavelength of radiation being under study. Thus in [8] the authors carried out experimental investigations of the backscattered optical resonant transition radiation emitted by 3-13 MeV electrons on a sub-micron grating produced by deformation of the conducting surface. The measurements were made for large observation angles and different radiation wavelengths. However, in view of complexity of the targets relief the experimental results did not gain any theoretical confirmation. Similar investigations presented in work [9] were carried out using the MAMI microtron in Mainz. In this work the authors also observed the resonant peaks within the optical range on the sub-micron optical grating. The transition radiation, diffraction radiation (the radiation emitted by a target during passage of a relativistic charged particle in the vicinity of the target edge) and the Smith –Purcell radiation (the radiation emitted by a relativistic charged particle moving over a periodic conducting target (grating) parallel to its surface) are traditionally considered. In reality, with angular divergence of an electron beam, particles moving at the angle to the target surface partially intersecting it (Fig.1).

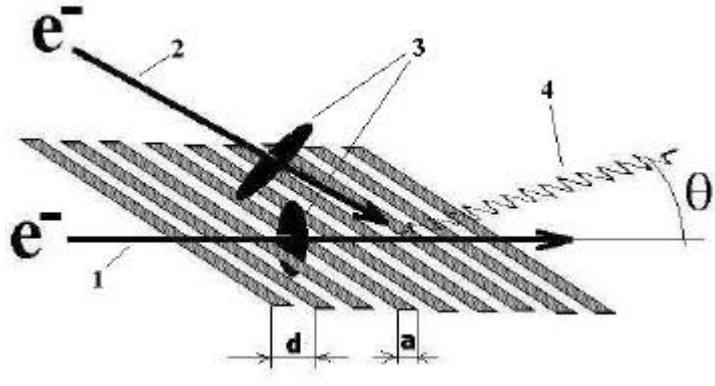

**Fig. 1.**

**To the polarization radiation concept.**

*1 is for the case of particle motion in the Smith-Purcell geometry,
2 is for the general case, 3 – is the particle electromagnetic field,
4 is the polarization radiation.*

All the above radiation types as well as their edge effects which cannot be described by traditional models may make their contributions. Thus, the study of resonant radiation at the inclined incidence of electrons on the periodic target is also high-priority task. It should be noted that such experimental studies have not been carried out earlier. The condition of resonance, which for the SMITH-Purcell geometry has the form $\lambda_k = \dfrac{d(\cos\theta - 1/\beta)}{k}$, is transformed to the expression $\lambda_k = \dfrac{d(\cos(\theta-\theta_0) - \cos\theta_0/\beta)}{k}$, where $k$ is an integer, $\lambda_k$ is radiation wavelength, $d$ is the target period, $\theta$ is the observation angle and $\theta_0$ is the inclination angle of the target to the direction of a relativistic particle velocity, $\beta = \sqrt{1-1/\gamma^2}$, $\gamma$ is the Lorenz factor of an electron.

Taking all of this into account we will consider a generalized "optical polarization radiation" (OPR) – the optical radiation of the target material during its polarization by an electric field of a relativistic charged particle. We next consider the polarization radiation from relativistic electrons, although the obtained results may be easily carried over to the case of other charged particles. We will also discuss this radiation from the periodic targets (see Fig.1) consisting of conducting strips separated by vacuum gaps or sputtered on a dielectric substrate. The effective transverse size of the field of a relativistic electron for its component with wavelength λ is equal to γλ, where γ is the Lorenz-factor of an electron. At sufficiently large γ this value may reach macroscopic sizes, thus enlarging in this way the area of the target polarization. In this case even at the inclined incidence of electrons on the target rather considerable number of strips may take part in radiation, which may substantially increase the resonant radiation intensity. On the other hand, a relativistic shift of the radiation frequency, if viewed at small angles to the direction of electron motion, enables one to obtain radiation from the macroscopic periodic targets (with period of hundreds of microns) in the optical range.

2. **Resonant optical polarization radiation of relativistic electrons. Theoretical model.**

An exact solution of Maxwell's equation for the problem, when a relativistic electron is moving in the vicinity of an infinitesimally thin, ideally conducting semi-infinity plane [5], is taken as a basis for the theoretical model to calculate radiation from the periodic targets described above. On the basis of this solution, the expressions for the fields $\vec{E}_{DR}$ the diffraction radiation from such a target were derived in [10]. We will consider $\vec{E}_{DR}(h)$ in such a form when it is assumed that $\vec{E}_{DR}(h)$ already contains the phase factor

$$e^{i\frac{2p}{l}h(\cos(q-q_0)-b\cos q_0)}, \text{ and the factor } \frac{e^{i\frac{2p}{l}R}}{R} \text{ is omitted.}$$

Here $h$ is the distance from the edge of the semi-infinity plane (SP) to the point of electrons intersecting the extension of this semi-infinity plane, $R$ is the distance from the target to observation point.

Then, using Babinet's principle we may write down the expression for the general case of an electron passage through the SP or in its vicinity

$$\vec{E}_{SS}(h) = \begin{cases} \vec{E}_{DR}(h) & h \notin SP \\ \vec{E}_{TR} - \vec{E}_{DR}(h) & h \in SP \end{cases},$$

where $\vec{E}_{TR}$ is the field of back-scattered transition radiation for the case when a relativistic electron is intersecting an infinitesimally thin, ideally conducting plane [11]. Now we can write down the expression for the field of radiation from a strip when an electron is intersecting it or is moving in its vicinity

$$\vec{E}_{Strip}(h) = \vec{E}_{SS}(h-a/2) - \vec{E}_{SS}(h+a/2).$$

Here $h=0$ corresponds to the electron passage through the middle of a strip, $a$ is the strip width. At last, the expression for the field of radiation from the periodic target consisting of $N$ strips with period $d$ will be of the form

$$\vec{E}_{Grat}(h) = \sum_{j=0}^{N-1} \vec{E}_{Strip}(h - j \cdot d).$$

After averaging over the impact parameter we will obtain

$$\vec{E}_{Average} = \frac{1}{(N+1)d} \int_{h=-d}^{N \cdot d} \vec{E}_{Grat\,(h)}(h) \, dh.$$

Now we may express the spectral and angular density of radiation

$$\frac{d^2 W}{d\boldsymbol{w} \cdot d\Omega} = 4\boldsymbol{p}^2 \left| \vec{E}_{Average} \right|^2.$$

Recall that the discussed model was obtained in the approximation of the infinitesimally thin, ideally conducting target. Further, the calculations were made in the approximation of small observation angles ($\boldsymbol{q}_0 \ll 1$) and a small angle of the inclined target with the electron beam directory ($1/\boldsymbol{g} \ll 1$). It is natural that the targets of finite thickness and real electric parameters were used in the experiment, although they were maximum approximated to the ideal ones. The same model was used for the case when the strips were sputtered on a ceramics substrate. The substrate polarization was not taken into account. While calculating radiation for the experimental conditions, the spectral and angular density of radiation was integrated over the angles within the limits of the aperture used in the experiments.

There are two possible methods of experimental study of the radiation angular characteristics. In the first method the angular distribution of radiation is directly measured by moving the detector in the two directions which are orthogonal to the observation direction. Under condition of strong shielding of the detector from x-ray background this method involves considerable technical difficulties. The second method consists in measuring the radiation dependence on the angles of the target orientation with respect to the direction of electron motion and the radiation plane (orientation dependence). With a precision goniometer for turning the target, this method enables one to obtain the same information as in the first one but with simpler means of realization.

Using the described model the authors calculated angular and orientation dependencies of the radiation intensities for the orthogonal components of the electric field vector with the fixed radiation wavelength as well as the spectra of the radiation intensity for a certain target position.

Two approaches to investigation of the resonant optical polarization radiation of relativistic electrons in the grating have been realized experimentally.

1. The investigation of the optical Smith-Purcell radiation.
   In this experiment the spectra of radiation from relativistic electrons passing in the vicinity of the periodic target (having the Smith-Purcell geometry) and intersecting it have been measured.

2. The investigation of the resonant optical polarization radiation from an inclined periodic target. The orientation dependence of radiation from relativistic electrons passing through an inclined grating have been measured. In this approach the radiation in a narrow wavelength range have been registered.

### 3. Investigation of the optical Smith-Purcell radiation.

The experiment was carried out on an extracted beam of the microtron which is an injector of the Tomsk synchrotron "Sirius".

The main parameters of the microtron are the following:
1. The energy of accelerated electrons $E_0$=6.1 MeV
2. The average current of the electron beam is 2.4 μA
3. The transverse size of the electron beam is 300 ~1000 μm

A periodic structure consisting of metal strips sputtered on a ceramic substrate was used as a target. The geometry of the target setup is shown in Fig.2.

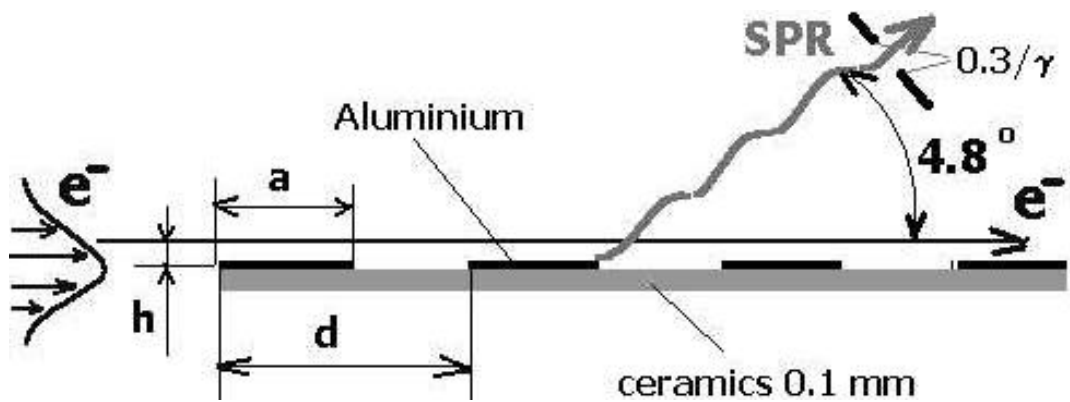

**Fig.2.**
**Geometry of the target setup.**

*h is the impact parameter, 0.3/**g** is the width of the opticalcollimator,*

***g** is the Lorenz factor, SPR is the Smith-Purcell radiation.*

In this experiment $d=0.1$ mm, $a=0.05$ mm. In view of the fact that the effective impact parameter $h_{eff}=\gamma\lambda \gg 10$ mm (here $\gamma$ is the Lorenz factor of an electron, $\lambda$ is the wavelength of the radiation under study) is much smaller than the electron beam size, the target was set up in the position of the maximum beam intensity. In this case a part of electrons intersected the target and was scattered by the atoms of the target.

Using the described model one may calculate the dependence of the radiation yield per unit of the solid angle on the radiation wavelength $\lambda$ and the angle of observation $\theta$ (Fig. 3).

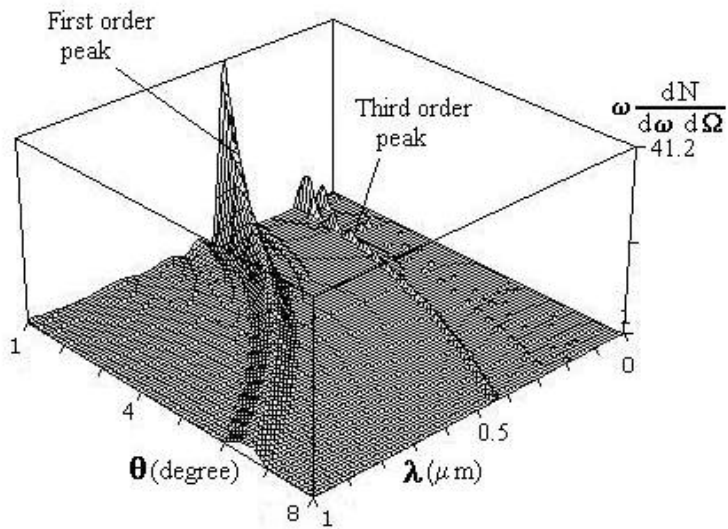

**Fig.3.**

*Spectral angular distribution of the Smith-Purcell radiation*

It is evident from the figure that one can observe a resonant first-order peak in the visible part of the spectrum ($\lambda \approx 0.7$ µm) at the observation angle $\approx 5°$. Fig. 4 shows the calculated intensity spectrum of radiation into the real solid angle for $\theta=4.8°$.

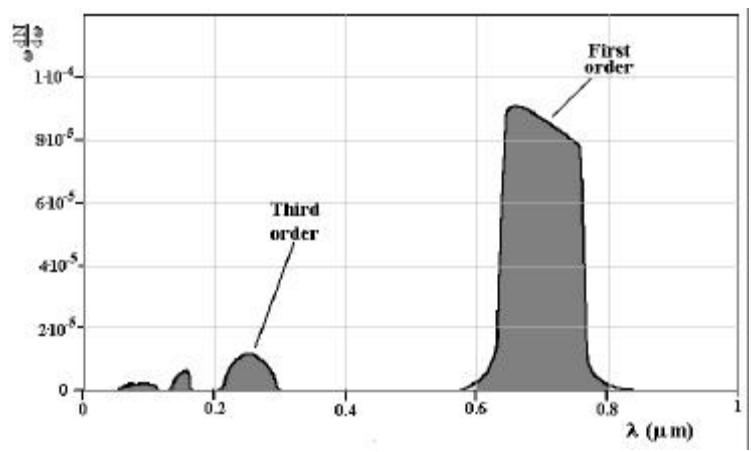

**Fig.4.**

An experimental setup for the measuring this spectrum is shown in Fig. 5.

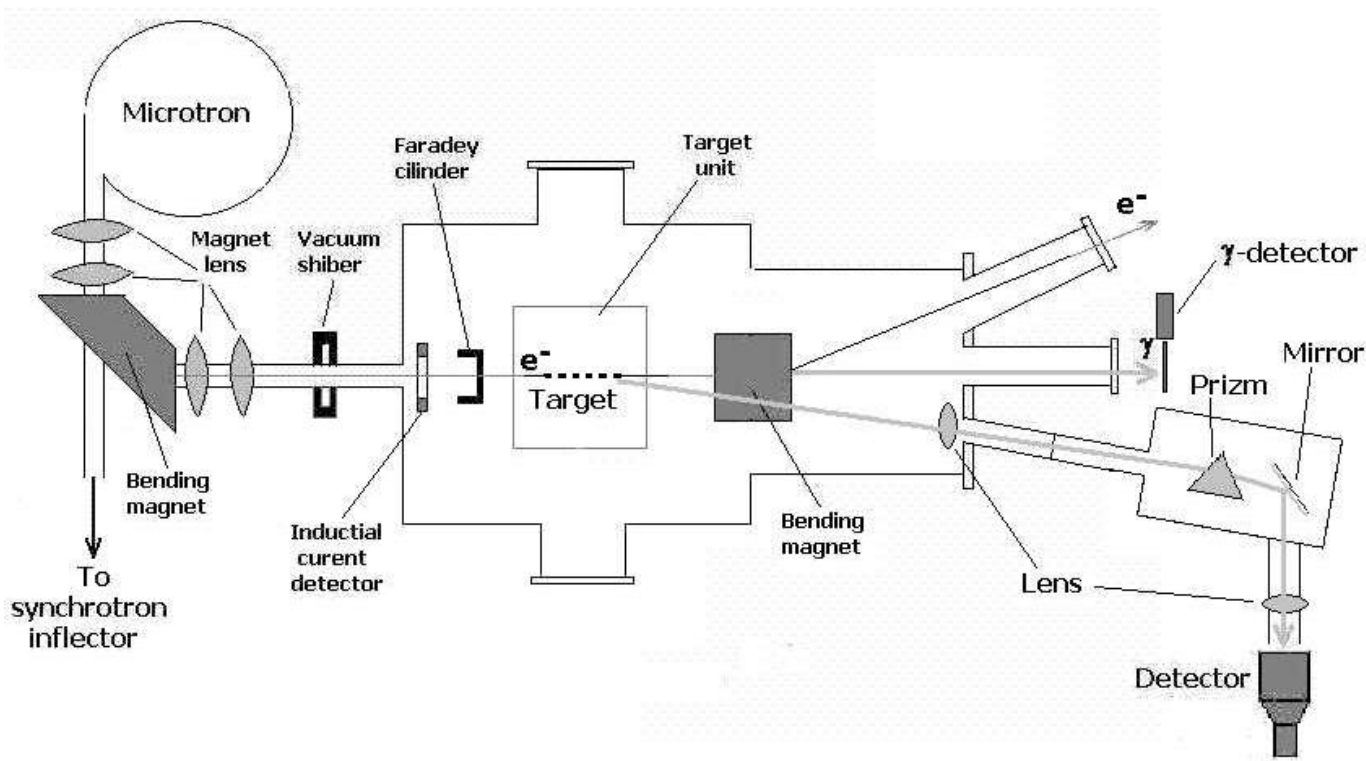

**Fig.5.**

A profile of the accelerator electron beam was measured using the bremsstrahlung detector with the vertical motion of the target. Optical radiation from the target was collimated by a slit collimator $0.3/\gamma$ width. An optical high-dispersion prism was performed to measure the spectrum. This fact enables us to resolve the spectrum into an angular distribution over the wavelength. The variation of the wavelength in the spectrum part being measured was achieved by the turn of the prism. The radiation was registered using a broadband photo-multiplying tube (PMT). Two optical lenses formed the radiation from the target. The first lens formed a parallel beam incident on the prism. The second lens focused the beam on the PMT. The resolution of the spectrometer was defined by the width of the slit collimator on the PMT and in the experiment it made up $\approx 20$ nm. The energy calibration of the spectrometer (Fig. 6) was made over the radiation from four light diodes with known wavelength.

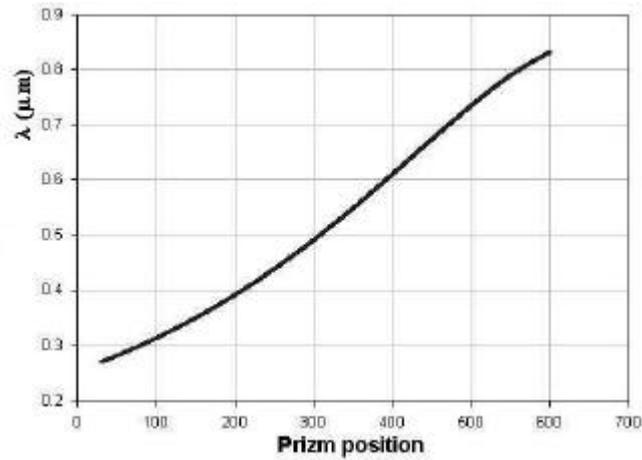

**Fig.6.**

*The wavelength registered by the spectrometer v.s. the angular position of the prism in units of the step monitor.*

The spectrum of radiation from the target was measured for two components of the polarization vector (for two components of the vector of radiation electric field: vertical and horizontal). For this purpose an optical polarizer was set at the optical path. To prevent the effect of the spectrometer time drift the polarizer was installed in two positions differing by the $90^o$ angle at each point under measurement. Also measurements of the x-ray background were taken at each point. For this purpose the optical path was cut by a light opaque blind. To measure the spectral efficiency of the whole spectrometric path we used a familiar spectrum of the transition radiation from a thin aluminum layer sputtered on a silicon substrate 0.43 mm thick. The spectral efficiency of the spectrometric path measured to the accuracy of an constant factor is shown in Fig. 7.

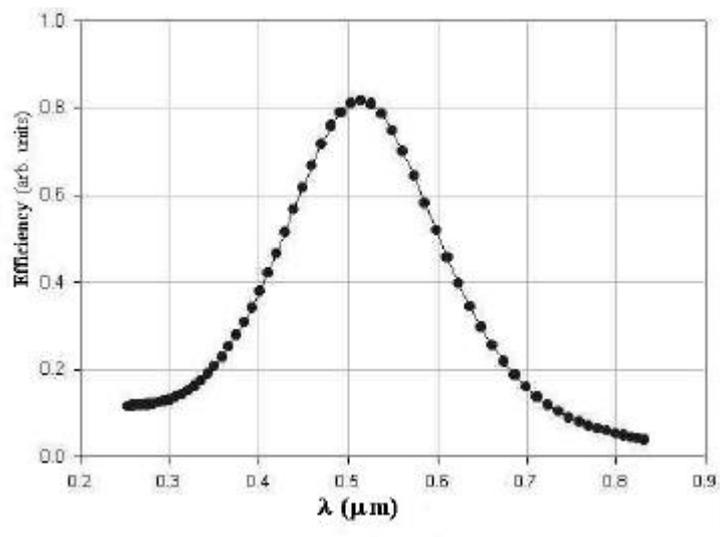

**Fig.7.**

By means of the described experimental setup the Smith-Purcell radiation spectrum was measured in the optical range for the two components of the radiation polarization (Fig. 8).

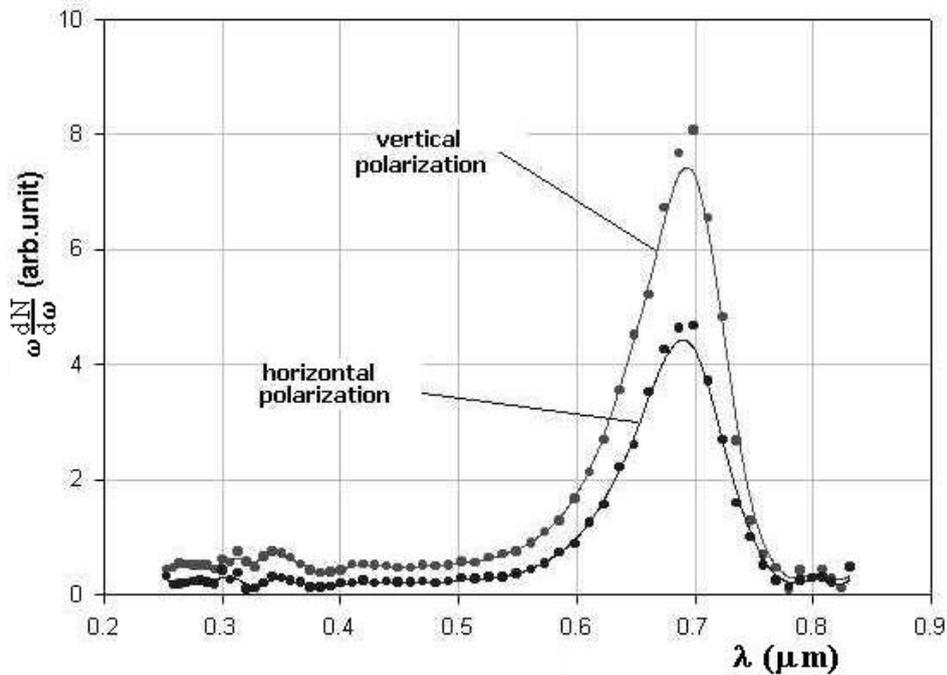

**Fig.8.**

*Optical spectrum for two components of the Smith-Purcell radiation polarization.*
*Solid curves show smoothed experimental spectra.*

The spectrum is given as corrected for the efficiency of the spectrometric path. Hereinafter the measured values in the experimental dependencies are denoted by dots and the curves smoothing data – by solid lines. In the figure one can see a pronounced first-order peak coinciding with the calculated one in its wavelength (see Fig. 4) as well as certain indications of a third-order resonance.

There areo even-order resonances in this geometry. Theoretically, in the Smith-Purcell radiation the horizontal (in the described geometry) polarization component is negligible. The presence of considerable contribution of this component in our experiment is explained by the resonant radiation of electrons scattered by the material of a ceramic substrate of the target. Unfortunately, this experiment did not allow us to obtain the absolute intensity of the radiation under study. The excess of the radiation intensity in the resonant peak over the intensity of measured transition radiation into the same aperture is $\approx 14$. The calculated value of this excess is equal to 12.5. The difference may be explained by a different shape of the theoretical and calculated resonant peaks as well as by an error in correcting for the multiple electron scattering in the target while the intensity of the transition radiation was being measured.

## 4. Investigation of resonance optical polarization radiation from an inclined periodic target.

This section is concerned with the results of investigating the orientation dependencies of the resonant radiation. The periodic conducting target with vacuum gaps between strips was used. The geometry of the target setup is shown in Fig. 9.

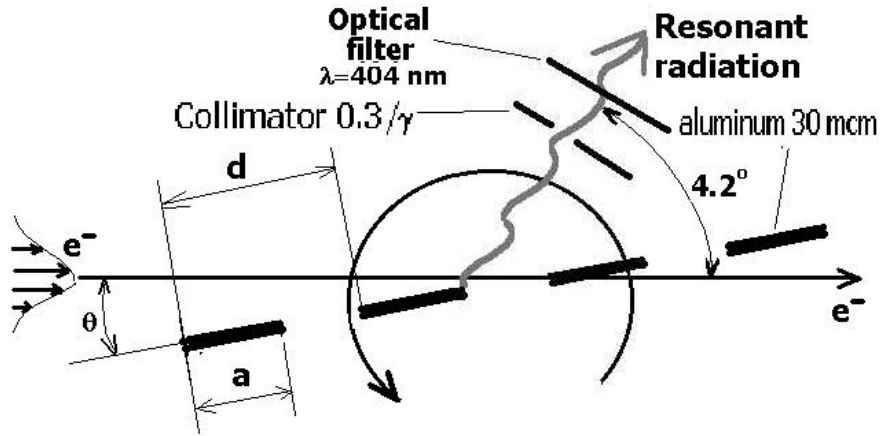

**Fig.9.**

*Geometry of the target setup on the path of electron beam.*
*d = 20 **m**m, a = 100 **m**m.*

The optical radiation under study was also registered by the PMT with low sensitivity to X-ray radiation. By means of the described model (1) one can calculate two-dimensional orientation dependence of this target radiation intensity on the angle of the target orientation in the reflection plane as well as on the orientation plane which is orthogonal to the electron beam direction (Fig. 10).

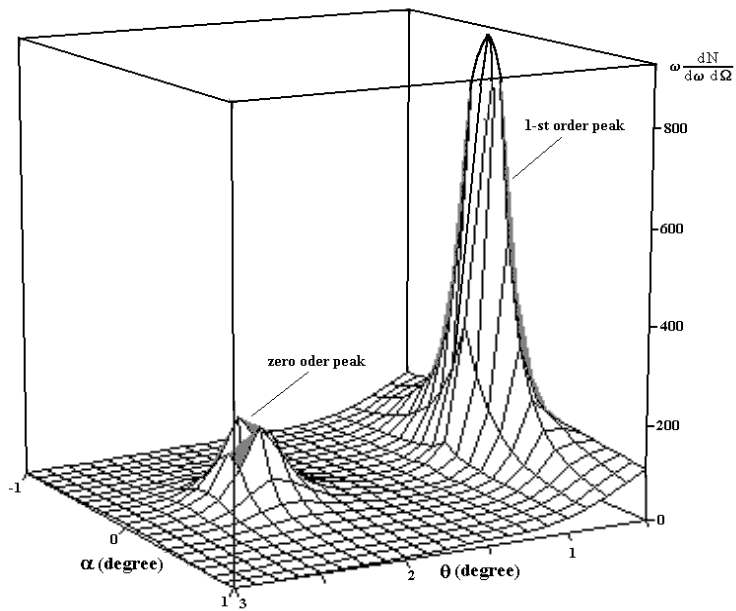

**Fig.10.**

*Calculated two-dimensional orientation dependence of the resonant radiation intensity.*

It is seen from the figures that one may expect a considerable contribution of radiation intensity in the area of the resonance peak as compared to the intensity in the zeroth reflex area.

Fig. 11 shows the calculated orientation dependence of the intensity of radiation with the 404 nm wavelength on the target inclination angle in the reflection plane into the experimental aperture.

A basic scheme of the experimental setup is shown in Fig. 12.

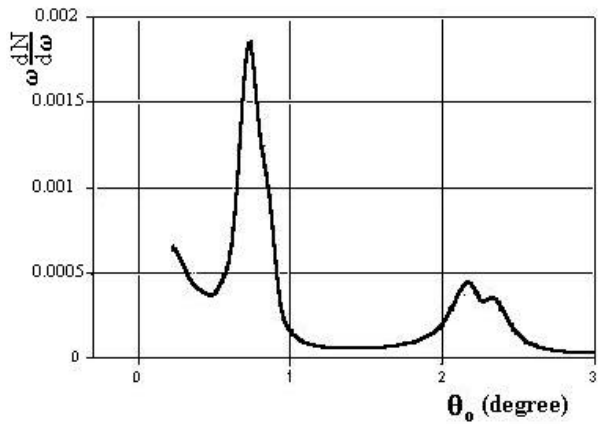

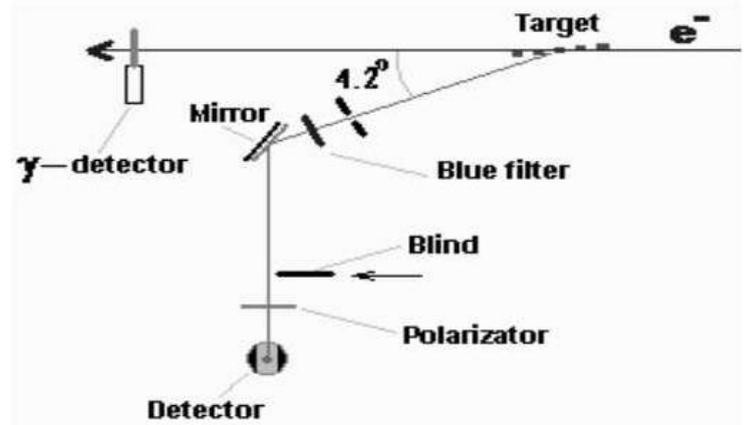

**Fig.11.**                                **Fig.12.**

The optical radiation from the target was collimated by a slit collimator having $0.3/\gamma \times 2/\gamma$ angular size. Radiation from the target was measured in the same way as in the previous experiment, for the two components of the polarization vector. For this purpose an optical polarizer was installed in the optical path, which was placed at each point under measurement in the two positions differing by an angle of $90^o$. As in the previous experiment, the X-ray background was measured at each point. An optical filter set up in the radiation path cut out a narrow spectral band $\Delta l \approx 20$ nm wide at the average wavelength $\lambda = 404$ nm. To control the experimental setup and the methods, the orientation dependence of the back-scattered transition radiation from a solid conducting target was measured (see Fig.13).

Comparison of the experimental dependence with the theoretical one calculated for the radiation into real aperture by the formula from [7] testifies to satisfactory potentialities of the setup and the methods (Fig. 14).

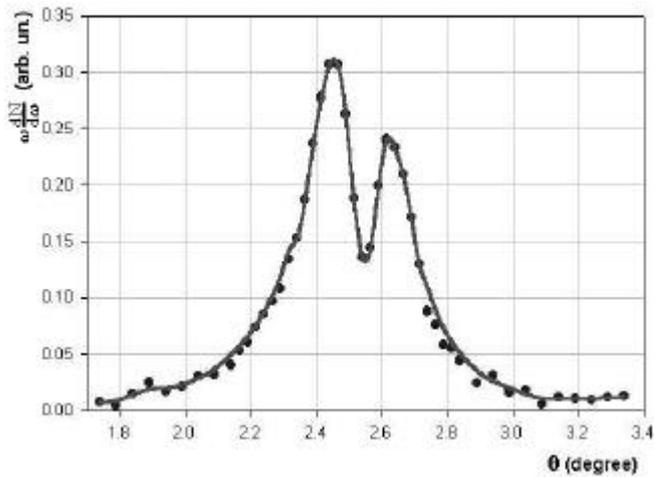
**Fig.13.**

*Experimental orientation dependence of back-scattered transition radiation from a solid conducting target.*

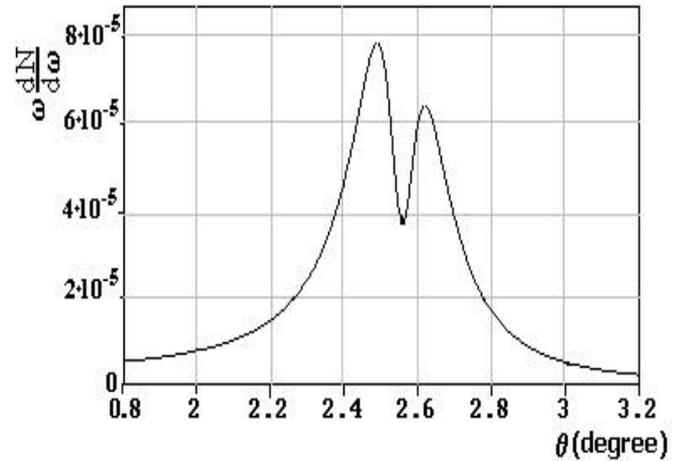
**Fig.14.**

*Theoretical orientation dependence of back-scattered transition radiation from a solid conducting target into the angular aperture 0.3/**g** x 2/**g***

In the experiment the orientation dependences of radiation from the periodic target in the reflection plane were studied. Fig.15 demonstrates the experimental orientation dependence of intensity of radiation from the periodic target in the absence of an optical filter and a polarizer.

In this figure one can see a very weak display of the first-order resonance peak at the angle è=0.7°. The position of the peak corresponds to the maximum spectral efficiency of PMT. Inclusion of the polarizer into measuring (Fig.16) exposes a polarization structure of the zeroth reflex (2~2.5°) but does not reveal the dependence in the area of the resonance peak.

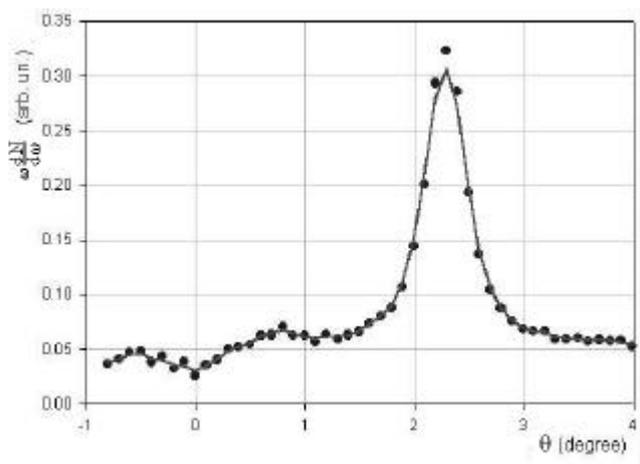
**Fig.15.**

*Experimental orientation dependence of the intensity of radiation from the periodic target in the absence of an optical filter and a polarizer.*

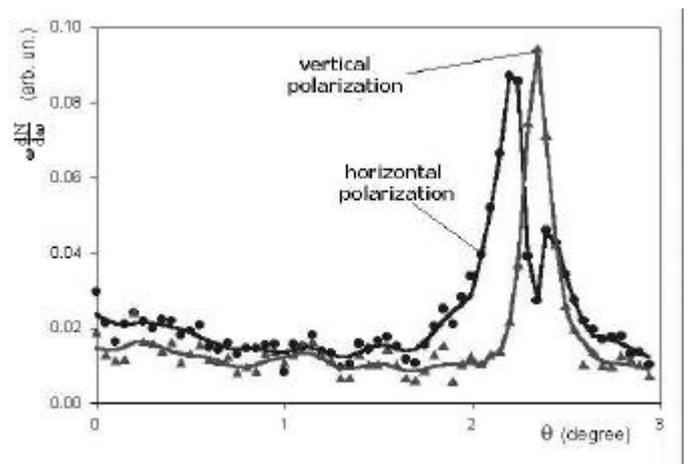
**Fig.16.**

*Experimental orientation dependence of the intensity of radiation from the periodic target for two components of radiation polarization.*

Inclusion of the optical filter without polarizer into the optical path enables us to see the resonance peak more distinctly, although with a considerable loss in statistics (Fig.17).

Since the cause for weak display of the resonance peak might be insufficiently high quality of the surface of the aluminum target (difficulties in preparing the surface for the target with vacuum gaps) similar measurements were taken with the target consisting of golden strips sputtered on a ceramic substrate. The geometry of target setup is similar to that shown in Fig.9. Orientation dependence of radiation from this target with inclusion of a light filter without polarizer into the optical path is shown in Fig.18.

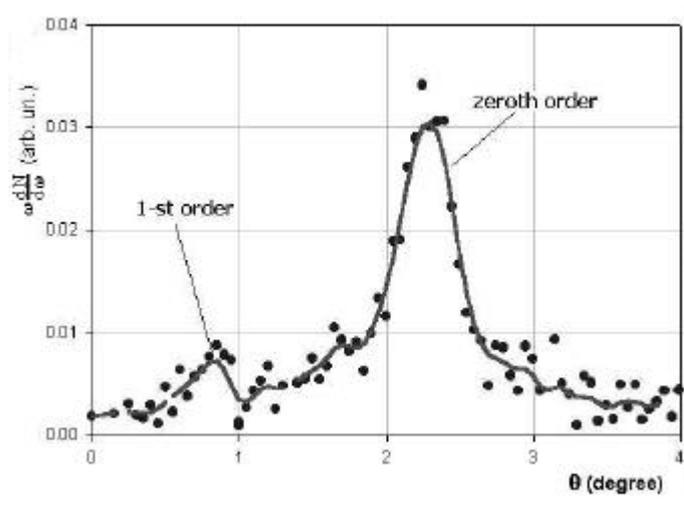

**Fig.17.**

*Experimental orientation dependence of the intensity of radiation through the optical filter of 404 nm wavelength.*

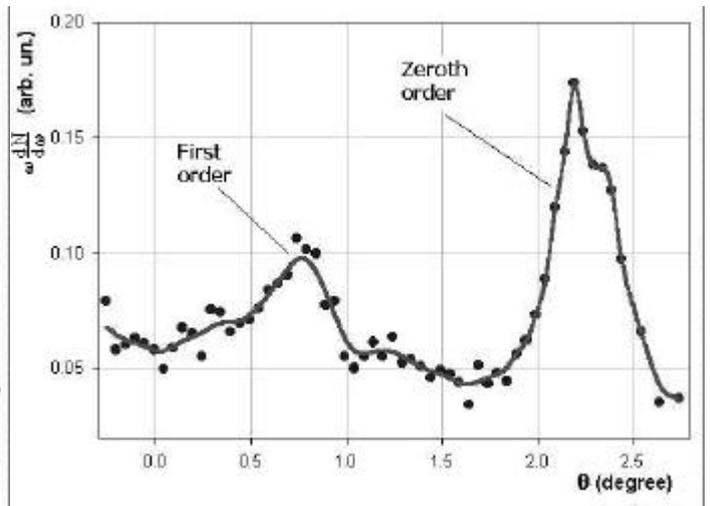

**Fig. 18.**

*Orientation dependence of radiation from the sputtered target with a light filter in the optical path.*

Here we can see much more pronounced display of the resonance although it is much smaller than theory predicts. In addition, even without a polarizer in the path one observes the structure in the zeroth reflex peak. Inclusion of a polarizer into the optical path reveals strong polarization of the resonant radiation in the reflection plane (normal to the strip direction) (Fig.19).

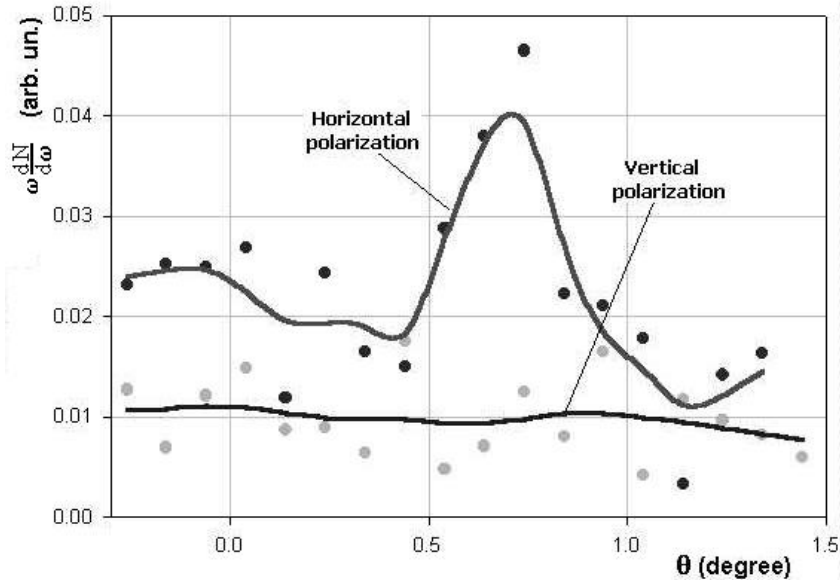

**Fig.19.**

*The experimental orientation dependence of radiation in the first-order resonance area with a light filter and a polarizer at the optical path.*

**Conclusion**

Despite their outward similarity (the investigation of resonances from a periodic target in the optical range) with experiments made in [8,9], the investigations carried out by the authors differ essentially from them in the properties of the object under study. The authors of [8,9] applied the optical gratings produced by deformation of the target surface, whereas in our case the targets consist of flat conducting strips either separated by vacuum gaps or sputtered on a dielectric substrate. In addition, the period of our targets is by 3 orders of magnitude higher than that of targets used in [8,9].

In our experiment two approaches were taken to the experimental investigations of the resonant polarization radiation from conducting periodic targets: spectral measurements of the Smith-Purcell radiation with electron energy 6.2 MeV and the measurements of the orientation characteristics of the resonant radiation with electron energy 200 MeV. In both cases the first-order peak of the resonant radiation was distinctly observed. The peak parameters in the Smith-Purcell geometry at low energy of relativistic electrons ($\gamma=12$) are satisfactory described by the theoretical model in the approximation of perfectly conducting, infinitesimally thin target with the use of Babinet's principle. However, at high energies of relativistic electrons ($\gamma=400$) the ratio of the resonant peak intensity of the radiation from the inclined target to that of transition radiation peak considerably differs from the theoretical ratio. The most probable cause for this discrepancy may be the fact that at large $\gamma$ the difference of the real target parameters from the ideal ones ( the finite target thickness and real electric characteristics of the target material) comes into play. Unfortunately, so far there is no approach to the theoretical solution of this problem for real target parameters. In the described experiments no qualitative deviations from the theoretical model (such as the Wood's anomaly in [8]) is found.

### Acknowledgments

This work was particularly supported by the RFBR (grant 99-02-16884) and by Program of state support of regional scientific high school politics of Russian Education Ministry and Atom-Ministry (grant 226).